\documentstyle[11pt,newpasp,twoside,epsf]{article}
\def\thetae{\theta_{\rm E}}
\def\te{t_{\rm E}}
\def\dos{D_{os}}
\def\dls{D_{ls}}
\def\dol{D_{ol}}
\def\drel{D}

\def\au{\rm AU}
\def\mjup{M_{\rm J}}

\def\msun{M_\odot}
\def\muas{\mu{\rm as}}
\def\mp{M_p}
\def\re{r_{\rm E}}

\def\edcomment#1{\iffalse\marginpar{\raggedright\sl#1\/}\else\relax\fi}
\reversemarginpar

\begin{document}
\title{Microlensing Searches for Extrasolar Planets: Current Status and Future Prospects}
  \author{B. Scott Gaudi}
\affil{Hubble Fellow, School of Natural Sciences, Institute for Advanced Study, Einstein Drive, Princeton, NJ, 08540}

\begin{abstract}
I review results from, and future prospects for, microlensing searches for
extrasolar planets.  Analyses of well-sampled microlensing light
curves by several collaborations have demonstrated that current
searches are sensitive to $\ga 1\mjup$ planets with few AU
separations from M dwarfs in the Galactic bulge.  To date, however, no
unambiguous planetary detections have been made.  Detailed analysis has
shown that this null result implies that $<33\%$ of typical stars
(i.e.\ M dwarfs) in the Galactic bulge have Jupiter-mass
companions with separations between $1.5$ and $4~\au$, and 
$<45\%$ have $3\mjup$ companions between $1$ and $7~\au$.
The recent dramatic increase in the number of alerts per year will
allow ongoing microlensing searches to probe companion fractions of a
few percent within a few years.  
\end{abstract}

\section{Introduction}

Ultra-precise radial velocity (RV) surveys have revealed over 100
planetary companions to nearby FGKM main-sequence stars.\footnote{See
http://cfa-www.harvard.edu/planets/catalog.html for a catalog and
discovery references}  The minimum masses $\mp$ and semi-major axes $a$ 
of these companions are shown in Figure 3.  The number of
known planetary companions is now reaching the point where robust
statistical inferences about trends within the sample can be made, and
thus the study of extrasolar planets is in transition from the
discovery phase to the characterization phase.

The `classical' methods of detecting extrasolar planets (RV,
astrometry, transits, direct detection) are generally complementary to
each other both in the range of semi-major axes $a$ they probe (see
Fig.\ 3), and in the parameters they measure.  They also
generally suffer from the same set of drawbacks.  First, because they
rely on light from either the parent star or the planet itself, they
are generally limited to nearby systems.  Second, they are
not currently sensitive to very low-mass planets.  For example, the
systematic floor of RV surveys is thought to be $\sim 1~{\rm m~s^{-1}}$.  
This implies that Earth,
Uranus, and Neptune analogs are probably inaccessible to RV surveys.
Finally, they generally require that the system be monitored for
at least one full period of the companion.  Thus, although RV surveys have
been ongoing for over a decade, they are only now
becoming sensitive to Jupiter-analogs (Marcy et~al.\ 2003).

Microlensing is an alternative method of detecting
planetary companions that overcomes many of the difficulties inherent
in the classical methods.  Mao \& Paczy\'nski (1991) first proposed
that microlensing could be used to detect planets; their ideas were
subsequently expanded on by Gould \& Loeb (1992).  Soon after these first
two seminal theoretical papers, several microlensing planet searches were
initiated (Pratt et~al.\ 1995, Albrow et~al.\ 2000), and microlensing
searches have now been ongoing, in some form, for nearly a decade.  Here I review the
basic theoretical concepts behind planetary microlensing, and describe how
microlensing planet searches work in practice.  I then give an
overview of what the analyses of actual datasets have taught us
about the practicality of the method itself, and about planetary companions in general.
Finally, I briefly speculate on future prospects for microlensing planet
searches.

\section{Microlensing and Planets}

Toward the Galactic bulge, microlensing occurs when a compact object,
typically a low-mass star, passes close to our line-of-sight to a more
distant star.  The lens splits the source
into two images separated by $\sim
2\thetae$, where
\begin{equation}
\thetae\equiv \sqrt{{{4 G M}\over \drel c^2} } = 300 \muas
\left(\frac{M}{0.3\msun}\right)^{1/2},
\label{eqn:thetae}
\end{equation}
is the Einstein ring radius of the lens, $\drel$ is defined by,
$\drel\equiv \dos\dol/\dls$, and $\dos$, $\dol$, and $\dls$ are the
distances between the observer-source, observer-lens, and lens-source,
respectively.  Thus for typical microlenses toward the bulge, the
images are unresolved, and the only observable is the sum of the
magnification of the two images of the source, which depends (only) on
the angular separation between the lens and source in units of
$\thetae$.  Since observer, source, and lens are in relative
motion, this angular separation, and thus the magnification, will be a
function of time: a microlensing event.  Normal microlensing events
have a characteristic, symmetric, three-parameter form.  The
parameters are the time of maximum magnification, the impact
parameter, and the characteristic timescale of the event,
\begin{equation}
\te\equiv \frac{\thetae}{\mu}=20~{\rm
days}\left(\frac{M}{0.3\msun}\right)^{1/2},
\end{equation}
where $\mu$ is the relative lens-source proper motion.  The timescale
is the only parameter that carries any information about the lens.
Figure 1 illustrates the image geometry as a function of the angular
separation of the source, and the resultant microlensing lightcurve.

\begin{figure}[t]
\plotfiddle{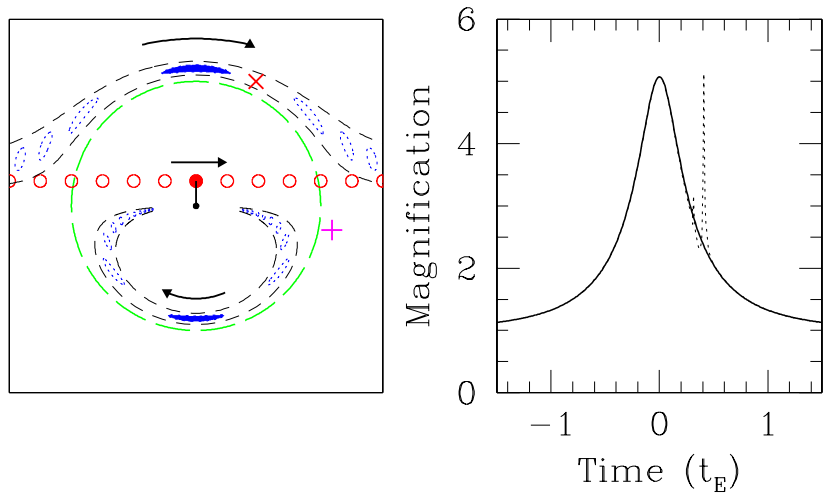}{0.5cm}{0}{70}{70}{-230}{-300}
\vskip6.0cm
\caption{
Left: The images (dotted ovals) are shown for several
different positions of the source (solid circles), along with the primary lens
(dot) and Einstein ring (long dashed circle).  If the
primary lens has a planet near the path of one of the images,
i.e. within the short-dashed lines, then the planet
will perturb the light from the source, creating a
deviation to the single lens light cure.
Right: The magnification as a function of
time is shown for the case of a single lens (solid) and accompanying planet
(dotted) located at the position of the X in the top panel.  
If the planet was located at the + instead, then there would be no detectable perturbation, and
the resulting light curve would be identical to the solid curve.
}
\end{figure}

If the primary lens has a planetary companion with projected position
near the path of one of the two images created in the primary event,
the planet will deflect the light from the image, creating a
perturbation atop the nominal single lens curve.  See Figure 1.  This
perturbation is the signature of a planetary companion to the primary
lens.  Three parameters describe the magnitude and shape of the
perturbation, and hence define the observables.  These
are the mass ratio $q$, the instantaneous projected separation $d$ between the planet and star, and the angle $\alpha$ of the source trajectory with respect 
to the planet-star axis.  The most interesting parameter
is $q$, which is related to and derived from the
duration of the perturbation, $t_p \simeq \sqrt{q} \te$,
which is $\simeq 1~{\rm day}$ for $\mp=\mjup$.
The orientation and phase of the planetary orbit, as well as $\thetae$, are all
generally unknown; thus $d$ gives only statistical information about
the semi-major axis $a$.  Since the planet must be near to one (or
both) of the two images in order to be detected, and the images are
near the Einstein ring radius while the source is
significantly magnified, microlensing is most sensitive to planets
with $d\sim 1$, which corresponds to $a \sim \dol \thetae \simeq
2\au \left({M}/{\msun}\right)^{1/2}$.  Finally, $\alpha$ is a randomly
distributed, geometric parameter, and is of no physical interest.

For fixed $q$ and $d$, only certain values of $\alpha$ will lead to
detectable perturbations (see Fig.\ 1). Therefore integration over
$\alpha$ defines a geometric detection probability.  This detection
probability is roughly $\sim A (\theta_p/\thetae) \simeq 15\%
(q/10^{-3})^{1/2}$, where $A$ is the instantaneous magnification and
$\theta_p\equiv \sqrt{q}\thetae$.  Thus the detection probability is
substantial, and high-magnification events are intrinsically more
sensitive to planetary companions (Griest \& Safizadeh 1998, 
Rattenbury et~al.\ 2003).

The advantages of microlensing are several.  First, no flux is needed
from either the lens or source, so systems with distances
of many kiloparsecs can be probed.
Second, it is sensitive to planets at separations $a=1-10~\au$
immediately, without monitoring for the entire orbital
period.  Finally, since the duration of the signal drops only as $\sqrt{q}$, it is
possible to extend the sensitivity to low masses.  The primary
disadvantages of microlensing are that follow-up of the detected
systems will be difficult due to their large distances, the
observations are non-repeatable, and the durations of the 
perturbations are short.  These last
two disadvantages combine to yield the requirement for successful
detection of planets via microlensing: nearly continuous, densely
sampled, and reasonably accurate photometry of the primary
microlensing events.

\section{Alerts and Follow-up}

The probability that a star toward the Galactic bulge will be
microlensed at any given time is of order $(10^{-6})$.  
Therefore, survey collaborations (EROS, Afonso et~al.\
2001; MOA, Bond et~al.\ 2001; OGLE, Udalski et~al.\ 2000) must monitor
several millions of stars at a time in order to detect ongoing events.
Practically, this means that several fields are monitored on a
once-per-night basis, and the photometry is generally not optimized
for any given microlensing event.  Therefore, current survey
photometry is generally not sufficient to detect planetary companions.
However, real-time data reduction enables these collaborations to issue
alerts, public notification of ongoing events.\footnote{For EROS alerts, see
http://www-dapnia.cea.fr/Phys/Spp/Experiences/EROS/alertes.html.}$^,\hskip-3.5pt$
\footnote{For MOA alerts, see
http://www.roe.ac.uk/$\sim$iab/alert/alert.html.}$^,\hskip-3.5pt$
\footnote{for OGLE alerts, see
http://www.astrouw.edu.pl/$\sim$ftp/ogle/ogle3/ews/ews.html.}  This
allows follow-up collaborations (EXPORT, Tsapras et~al.\ 2001; MPS,
Rhie et~al.\ 2000; PLANET, Albrow et~al.\ 2000) to monitor only those
events that are ongoing at any given time, and to optimize sampling
rates and exposures times to individual events.

An example of this two-tier procedure is shown in Figure 2.  On the left is the OGLE lightcurve for 
the alert OGLE-1998-BUL-14.  On the right is the PLANET photometry of
the same event (Albrow et~al.\ 2000).   While the OGLE dataset
is relatively sparse, the median sampling interval for the
PLANET data 
is about 1 hour, or $10^{-3}\te$, with
very few gaps greater than 1 day.  The $1\sigma$
scatter in $I$ over the peak of the event (where the sensitivity to
planets is the highest) is $1.5\%$.  The dense sampling and excellent
photometry means that the efficiency to detect massive
companions in this event should be quite high.  In fact, a rigorous
search for planetary companions, and quantification of the detection 
efficiency (Gaudi \& Sackett 2000), reveals that, although the dataset is indeed quite
sensitive to $\mjup$ planets at separations of $a=1-5~\au$, there is no evidence
for a companion.   Analyses of similar datasets of other events using a variety of 
techniques have clearly demonstrated that the requisite photometric accuracy and sampling needed to
robustly detect planetary perturbations is readily achievable (Albrow et~al.\ 2000, Rhie et~al.\ 2000, Bond et~al.\ 2002, Tsapras et~al.\ 2002). 
 Therefore microlensing is
not only sensitive to $\sim \mjup$ companions between $1-10~\au$ in principle, but also in practice.

\begin{figure}[t]
\plottwo{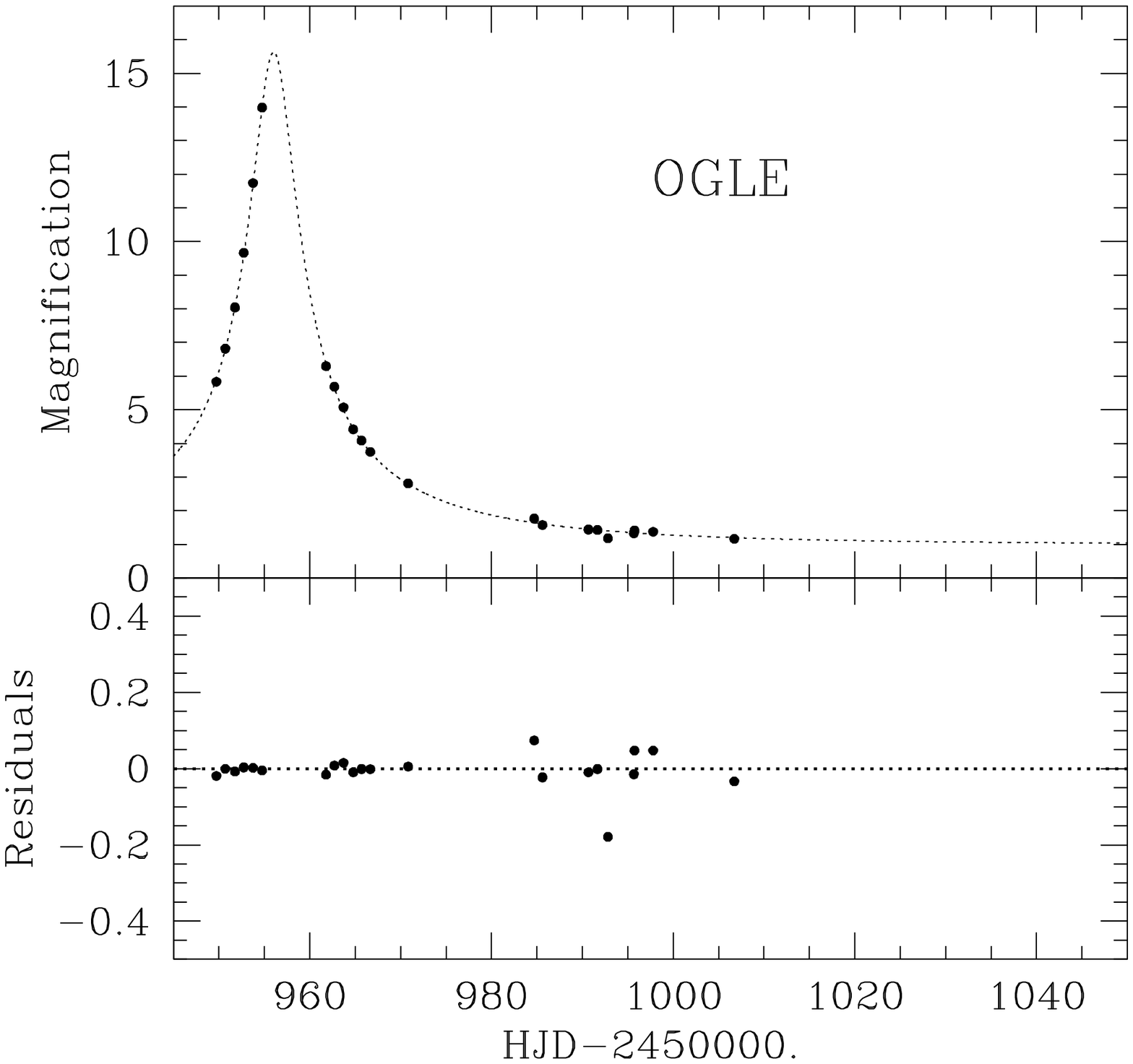}{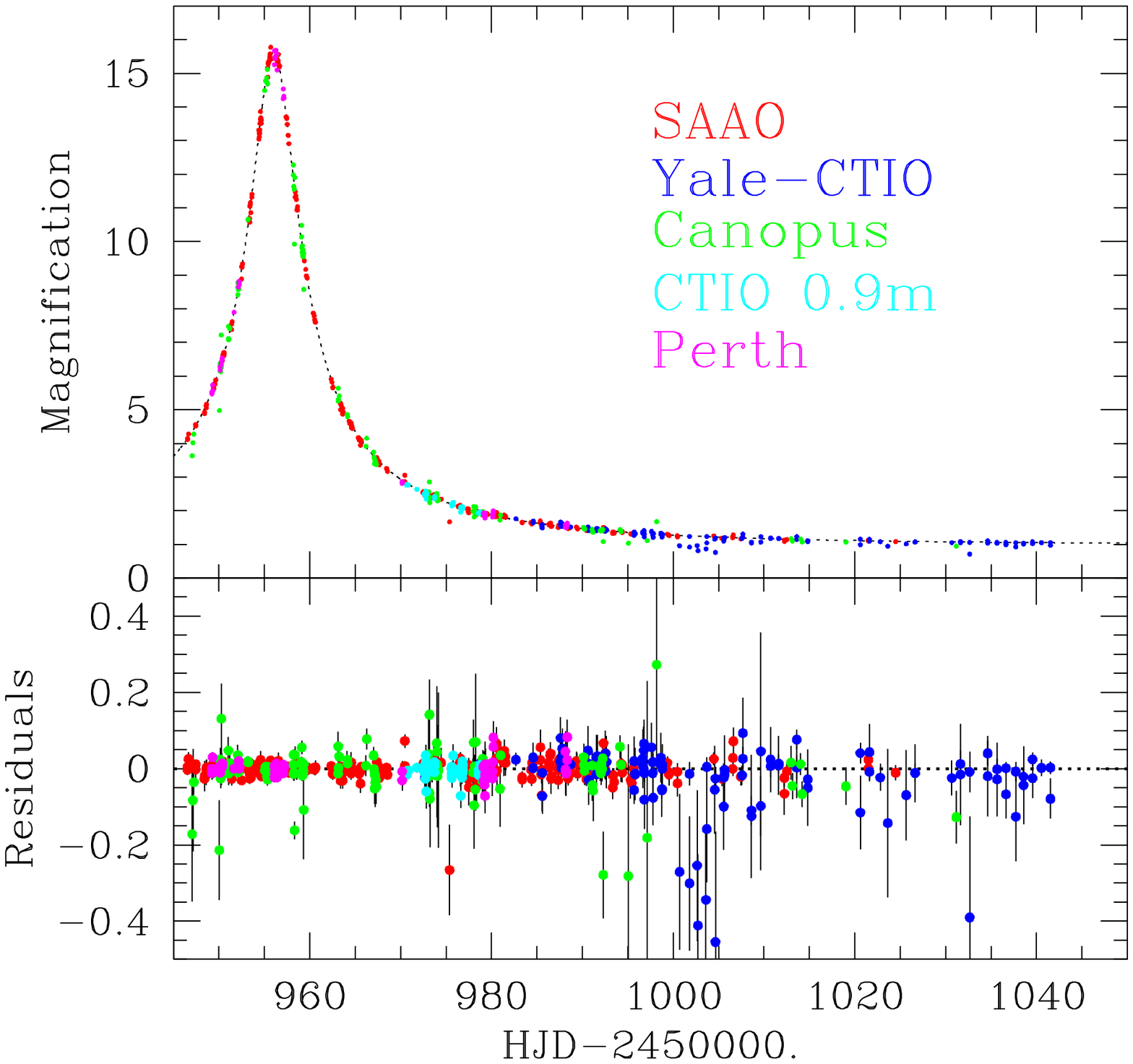}
\caption{
Left: The top panel shows the magnification as a function of time for 
OGLE data of microlensing event OGLE-1998-BUL-14.  The dashed line indicates the best-fit point-lens model.  The bottom panel
shows the residuals from the best-fit model.
Right: The top panel shows the PLANET data for the same event.
The bottom panel shows the residuals (Albrow et~al.\ 2000).}
\end{figure}

\section{Limits on Galactic Planets}

Analyses of datasets of individual microlensing events have
demonstrated that microlensing is a practical method to detect $\sim
\mjup$ planetary companions, and in some have cases ruled out
companions to individual lenses over a large range of separations.
However, a coherent, uniform analysis of a significant sample of events 
is necessary to draw conclusions about the population of companions to lenses
as a whole.   The first such study was performed by the PLANET collaboration, who
analyzed their complete database of photometric measurements of 
bulge microlensing events during the first five years of 
their campaign, from 1995 to 1999 (Gaudi et~al.\ 2002).  
After applying cuts to eliminate nearly
equal-mass binaries, poorly sampled events, and poorly-constrained events, a
well-defined sample of 43 events was analyzed.

\begin{figure}
\plottwo{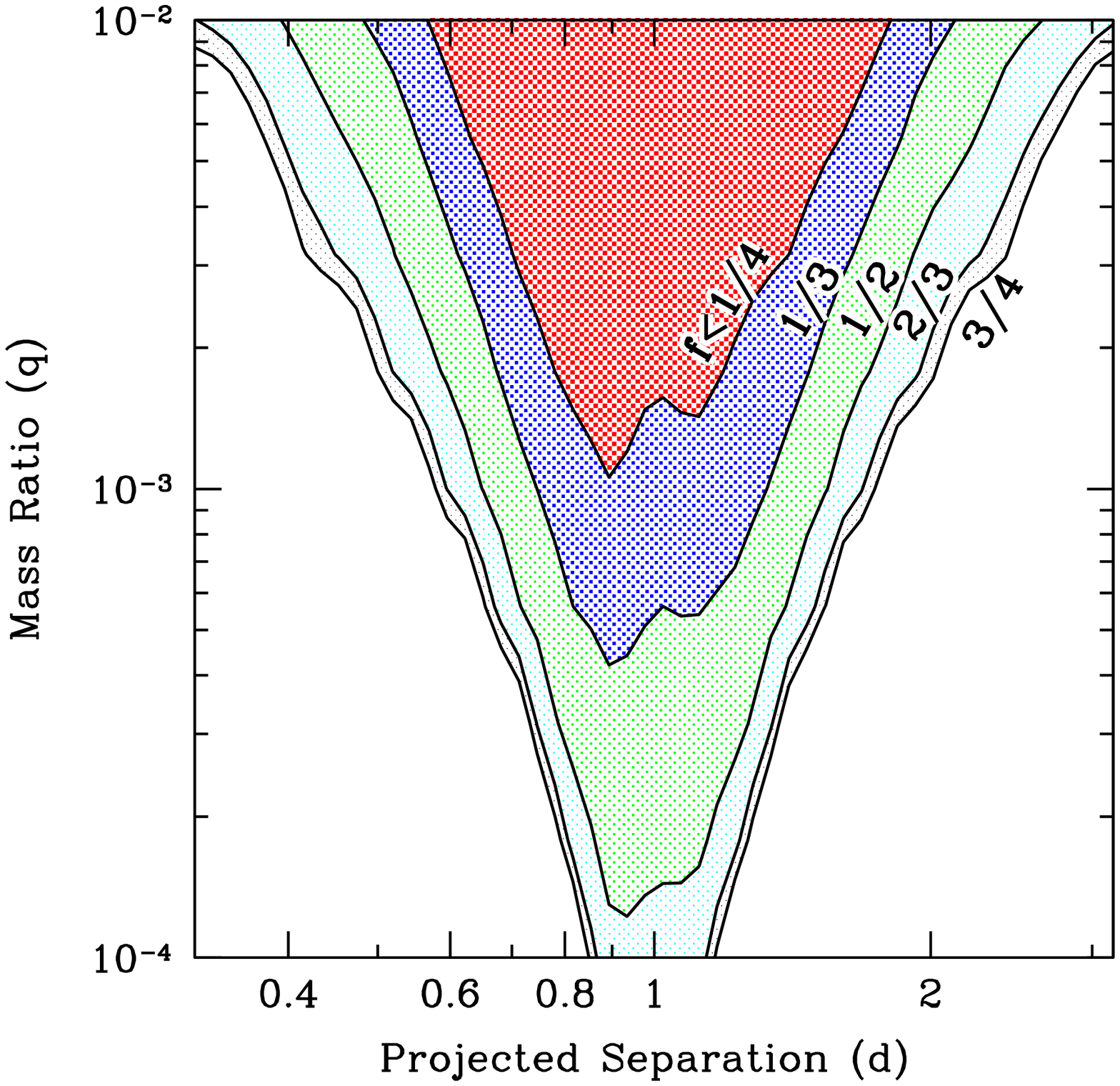}{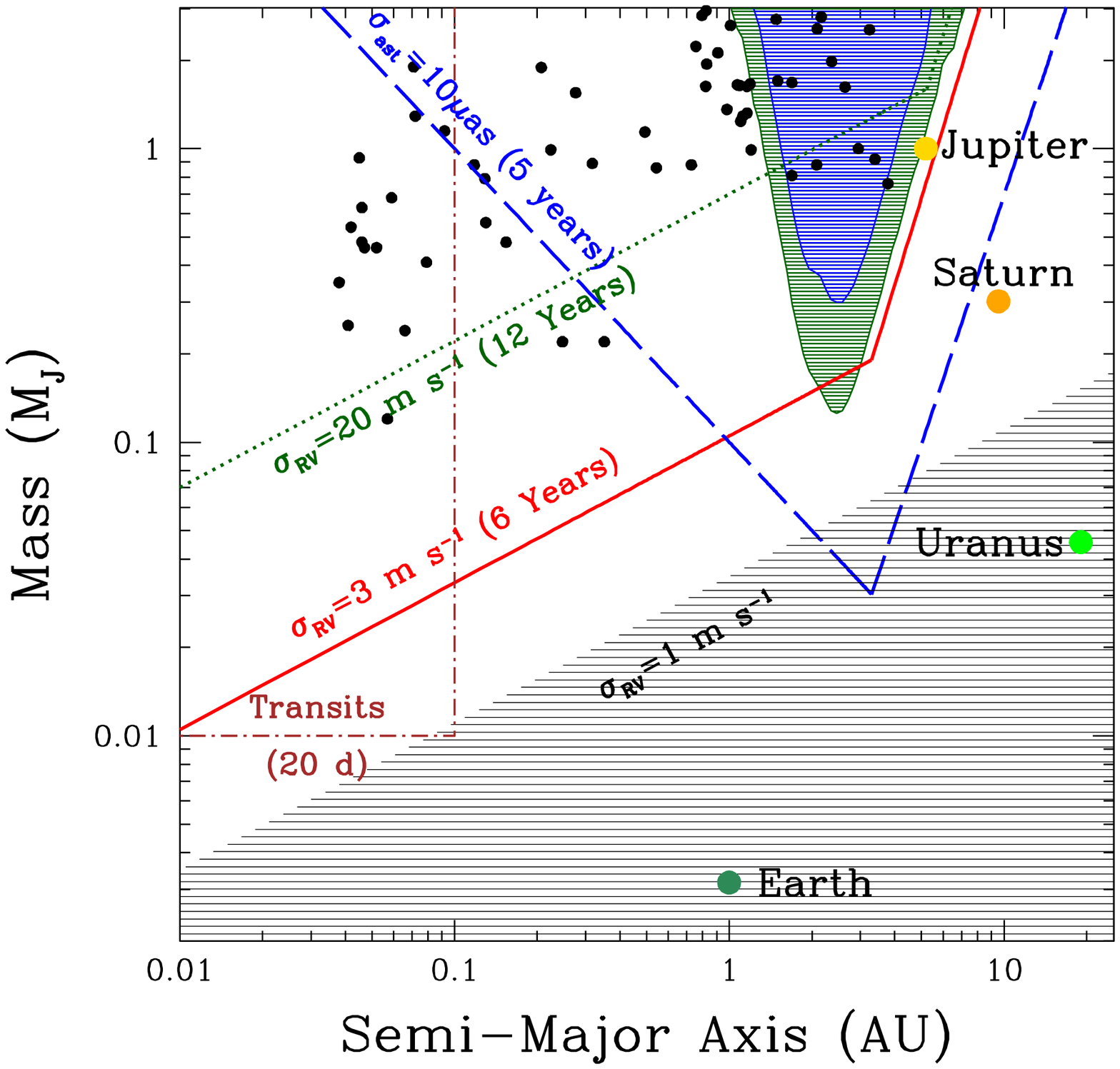}
\caption{
The left panel shows contours of the 95\% confidence upper limit 
to the fraction of lenses with a companion as a function of the
mass ratio $q$ and projected separation $d$ of the companion from the 
PLANET 5-Year analysis (Albrow et~al.\ 2001).  The right panel is a
census of planetary companions to solar-type
stars, as well as sensitivities of various detection methods.   
The exclusion region from the PLANET microlensing
survey is shown as the triangular hatched regions; $<45\%, <33\%$ (outer, inner) of M dwarfs in the bulge have companions in these regions.  }
\end{figure}

Each event was searched for planetary perturbations.  
There were no clear detections.  For each event, the detection
efficiency  was computed for a broad range of $q$ and $d$, using the method of Gaudi \& Sackett (2000). These efficiencies were then combined to produce an upper limit to the fraction $f(d,q)$
of lenses with a companion.  The $95\%$ confidence level (c.l.) upper limit to $f(d,q)$ from the PLANET five year analysis is shown in Figure 3.  

Although microlensing constraints are the most robust and least model-dependent
in the $(d,q)$-plane, this parameterization makes it difficult to compare with other methods, which are primarily sensitive to $\mp$ and $a$.  To convert from $(d,q)$ to $(a,\mp)$, it is necessary to specify
the typical mass $M$ and physical Einstein ring radius $\re=\dol\thetae$ of
the primaries, and also to marginalize over all possible orientations and orbital
phases (as $d$ is the instantaneous projected separation).  It is likely that the majority of the 
microlensing primaries are
M dwarfs in the Galactic bulge (Gould 2000, Albrow et~al.\ 2001). 
Thus $M\simeq 0.3\msun$ and $\re \simeq 2\au$.  With these values, 
$q=0.001\Rightarrow m_p=0.3\,\mjup$, and $d=1\Rightarrow
r_p=2\au$.  Figure 3 shows the resulting constraints in the $(a,\mp)$-plane.

The PLANET analysis demonstrates that $<33\%$ of M dwarfs in the
Galactic bulge have Jupiter-mass companions with
separations between $1.5$ and $4~\au$, and less than 45\% have
$3\mjup$ companions between $1$ and $7~\au$.  The range of
parameter space constrained by PLANET observations partially overlaps
that surveyed by RV studies; indeed many of the planets detected by RV
surveys fall within the PLANET 45\% exclusion region.  However, the
fractions of local stars with planets in this range is $\sim 5\%$, and
so the RV results are not in conflict with microlensing constraints.
Indeed, as RV and microlensing surveys are probing disjoint
populations, both in regards to the typical mass of the primaries
($M\sim \msun$ for RV versus $M\sim 0.3\msun$ for microlensing), and
the parent populations (local stars for RV versus bulge stars for
microlensing), a comparison between RV and microlensing results would
likely provide clues as to how planet formation and survival depends on the
primary mass and local environment.  As I discuss in the next section,
although microlensing constraints have not yet reached the fractions
currently probed by RV surveys, continued monitoring should allow a
meaningful comparison between the two methods in the next few years.

\section{Future Prospects}

In the immediate future, microlensing searches for extrasolar planets
are likely to continue with the current two-tier ``survey/follow-up'' system.
How can such searches be improved, and the rate at which statistics
are obtained increased?  Some improvement can be achieved by more
efficient monitoring of events and/or by employing image subtraction,
rather than PSF-fitting, photometry.  However, the most crucial factor
in the efficiency of the survey/follow-up system is the number and
quality of the alerts provided by the survey collaborations.  A large
number of alerts allows the follow-up collaborations to monitor only
those events that are the most promising and most efficient for the
detection of planetary perturbations, i.e.\ bright, high-magnification
events.  Previous to 2002, $\la 100$ alerts per year were provided
by the various survey collaborations.  However, the OGLE collaboration
recently upgraded its camera system (Udalski et~al.\ 2002), and an
extrapolation from the number of events alerted so far
this year suggests that $\sim 500$ per year can be expected.
Assuming photon-limited photometry, optimal observational
strategies, and that a fraction $f(q)$ of lenses have companions with
mass ratio $q$, and separations uniformly distributed in the range $d=0.5-2.5$, the rate ${\cal R}_{det}$ 
at which planetary detections can be expected is roughly,
\begin{equation}
{\cal R}_{det} \sim 1~{\rm yr^{-1}}
{\left(q \over 10^{-3}\right)}^{1/2}
{\left({\cal R}_{alert} \over 500~{\rm yr^{-1}} \right)}^{1/2}
{\left(\frac{f(q)}{5\%}\right)}.\label{eqn:ndetscale}
\end{equation}
Thus current survey/follow-up microlensing searches should probe companion frequencies of a few percent over the next several years.  

Microlensing is unique among planet detection methods in that, to
zeroth order, the amplitude of the signal does not decrease as $q$
decreases. Therefore, very low-mass planets are in principle
detectable via microlensing.  The ultimate limit is set by the size of
the source star: if the angular size of the source $\theta_*$ is
larger than $\theta_p$, the signal will be `washed out' by the finite
size of the source.  For main-sequence sources, this limit is reached
at about $q \simeq 10^{-5}$, the mass ratio between an Earth mass planet
and the typical microlensing primary.  The duration and detection
probability both decrease as $\sqrt{q}$.  For $q=10^{-5}$, the
duration of the perturbation is $\sim 1~{\rm hr}$ and the detection
probability is a few percent (Bennett \& Rhie 1996)

Even if the photometric accuracies and sampling rates required
to detect the 
perturbations from Earth-mass planets are achieved, a naive
extrapolation of equation (3) suggests that it will
be quite difficult to build up reasonable statistics on Earth-mass
planets using the survey/follow-up method.  This is simply due to the
low detection probability and the finite total exposure time.
Ground-based detection of Earth-mass planets might be possible via a
`next-generation' ground-based microlensing planet search.  Such a
search dispenses with the survey/follow-up model.  Instead, it employs 
a network of 2m-class telescopes distributed throughout the Southern
hemisphere equipped with large-format CCDs that continually monitor a
few fields toward the bulge at extremely high cadence to
simultaneously detect {\it and} monitor events for planets.  

The primary difficulties with ground-based surveys for Earth-mass
planets are weather, which hampers the ability to robustly detect and
constrain the planetary perturbations, and atmospheric seeing,
which causes main-sequence
sources to be severely blended even
in exceptional conditions, thereby increasing the photometric
noise.  A space-based mission would overcome these difficulties.  The
Galactic Exoplanet Survey Telescope (GEST), a proposed 2m aperture
telescope with a $\sim 2$ square degree field-of-view would provide
interesting constraints on Earth-mass planets, including Earth
analogs (Bennett \& Rhie 2002).  The primary disadvantage of a space-based survey is the high cost.

Both next-generation ground-based surveys and space-based missions
would also be sensitive to very wide separation and free-floating
planets (Di Stefano \& Scalzo 1999a,b).  Microlensing is the only
method yet proposed to detect low-mass free-floating planets.

\acknowledgements

I would like to thank the SOC and LOC for organizing
an enjoyable and productive conference. 
Support for this work was provided by NASA through a Hubble Fellowship grant
from the Space Telescope Science Institute, which is operated by the
Association of Universities for Research in Astronomy, Inc., under
NASA contract NAS5-26555.

\end{document}